\documentclass[preprint,showpacs,rmp]{revtex4}
\usepackage{epsfig}

\begin{document}
\title{Shot noise in metallic double dot structures with
a negative differential conductance}

\author{V. Hung Nguyen and V. Lien Nguyen \footnote{Corresponding author,
E-mail: nvlien@iop.vast.ac.vn} } \affiliation{Theoretical Dept.,
Institute of Physics, VAST, P.O. Box 429 Bo Ho, Hanoi 10000,
Vietnam}
\author{Philippe Dollfus}
\affiliation{Institut d'$\acute{E}$lectronique Fondamentale (CNRS
UMR 8622), Universit$\acute{e}$ Paris XI, 91405 Orsay, France}

\begin{abstract}
The shot noise of current through a metallic double quantum dot
structure exhibiting negative differential conductance is studied.
We can exactly solve the master equation and derive an analytical
expression of the spectral density of current fluctuations as a
function of frequency in the first Coulomb staircase region. For a
large range of bias voltage the noise is calculated by Monte-Carlo
simulation. We show that the noise is always sub-Poissonian though
it is considerably enhanced in the negative differential
conductance regime.
\end{abstract}

\pacs{73.63.Kv, 73.40.Gk, 72.70.tm}

\maketitle

The shot noise (SN) is a striking consequence of charge
quantization and its study has become an emerging topic in physics
of nano-devices because measurement of the SN can reveal more
information on transport properties which are not available
through the conductance measurement alone [1,2]. In the case of
uncorrelated current the SN has the full (or Poissonian) value
$2eI$, where $e$ is the elementary charge and $I$ is the average
current. Deviations (either suppression or enhancement) from this
value are due to correlations in the motion of charge carriers.
The measure of these deviations is the Fano factor $F$ which is
defined as the ratio of the actual noise spectral density to the
full SN-value. There are mainly two kinds of correlation: the
Pauli exclusion principle and Coulomb interaction. While the
former correlation always causes a suppression of SN, the latter
may suppress or enhance the noise depending on the conduction
regime. The non-Poissonian noise has been most extensively studied
in double barrier resonant tunneling diodes, where the SN is
partially suppressed in the positive differential conductance
(PDC) regime and becomes super-Poissonian in the negative
differential conductance (NDC) regime [3]. The super-Poissonian
noise accompanied by an NDC has been also predicted and observed
in quantum dot devices [4,5]. For Coulomb blockade metallic
structures when the Coulomb interaction is extremely important and
manifested explicitly in the so-called charging energy, the SN has
been studied in a number of works focussing on the single-electron
transistor (SET) in PDC regime [6-8] (and refs. in [1]). It was
shown that the noise is generally suppressed with a magnitude
depending on device parameters and on the range of applied
voltage. Recently, we were able to derive an analytical expression
for current-voltage (I-V) characteristics and a condition for
observing an NDC in a Metallic Double Quantum Dot Structure
(MDQDS) in the first Coulomb staircase region [9]. The NDC has
been analyzed in detail with respect to device parameters as well
as to the temperature and the off-set charge. Then, it is natural
to raise a question about the SN-behavior in this device in an NDC
regime. We will show in the present letter that the noise though
considerably enhanced in NDC regime seems to be always
sub-Poissonian. This has been done by solving exactly the master
equation in the first Coulomb staircase region and by Monte-Carlo
(MC) simulation in a large range of bias voltage.

The equivalent circuit diagram of the structure studied is drawn
in the inset of Fig.1a. Within the framework of the Orthodox
theory the state $|i>$ of the system is described by the
probability $p(i,t)$ to have $n_i$ (and $m_i$) excess electrons in
the dot $D_1$ (and $D_2$). This probability obeys the master
equation which can be written in the matrix form as $d\hat p\left(
t \right)/dt = \hat M\hat p\left( t \right)$, where $\hat p\left(
t \right)$ is the matrix of elements $p(i,t)$ and $\hat M$ is the
evolution matrix of elements $M_{ij} = \Gamma (j \leftarrow i) -
\delta _{ij} \sum\nolimits_k {\Gamma (k \leftarrow i)}$, with
$\Gamma (j \leftarrow i)$ being the net transition rate from $|i>$
to $|j>$ (see [6,7] and formula (1) in ref.[9] for the model under
study). The statistically averaged current across the junction
$\nu$ ($\nu = l, m$ or $r$) is $\left\langle {I_\nu \left( t
\right)} \right\rangle  = e\sum\nolimits_i {\left[ {\Gamma_{\nu}^+
(i)  - \Gamma_{\nu}^- (i)} \right]p (i,t)}$, where
$\Gamma_{\nu}^{\pm }(i)$ is the tunneling rate through the
junction $\nu $ to the right(+)/left(-) at the state $|i>$. For a
stationary state the current is $t$-independent and the total
external current is $I = \left\langle {I_\nu  } \right\rangle$ for
any of $\nu$. However, if the state is not stationary, the charge
accumulated in the dots is time dependent and $\left\langle {I
(t)} \right\rangle$ is a weighted average of $\left\langle {I_\nu
(t)} \right\rangle$ as $\left\langle {I\left( t \right)}
\right\rangle = \sum\nolimits_\nu  {g_\nu \left\langle {I_\nu
\left( t \right)} \right\rangle }$, where $g_l = C_r C_m /
\Sigma_c , \ g_m = C_l C_r / \Sigma_c , \ g_r = C_l C_m /
\Sigma_c$ and $\Sigma_c = C_l C_m + C_m C_r + C_l C_r$. We will be
interested in the SN for $I$, which is the external current and
experimentally measurable [10], as well as for partial currents
$I_{\nu}$.

Korotkov [7] suggested in detail the procedure of calculating the
noise spectrum of a correlated tunneling current in a SET.
Extending this procedure for the MDQDS of interest we have
\begin{eqnarray}
\begin{array}{l}
 S_{\nu \nu } \left( \omega  \right) = 2A_\nu   + 4e^2 \sum\limits_{ij} {\left[ {\Gamma _\nu ^ +  \left( i \right) - \Gamma _\nu ^ -  \left( i \right)} \right]B_{ij} \left[ {\Gamma _\nu ^ +  \left( {j\left| {\nu ^ -  } \right.} \right)p_{st} \left( {j\left| {\nu ^ -  } \right.} \right) - \Gamma _\nu ^ -  \left( {j\left| {\nu ^ +  } \right.} \right)p_{st} \left( {j\left| {\nu ^ +  } \right.} \right)} \right]},  \\
 S_{II} \left( \omega  \right) = 2\sum\limits_\nu  {g_\nu ^2 A_\nu  }  + 4e^2 \sum\limits_{\nu \mu} {\sum\limits_{ij} {g_\nu  g_\mu  \left[ {\Gamma _\nu ^ +  \left( i \right) - \Gamma _\nu ^ -  \left( i \right)} \right]B_{ij} }}  \\
 \ \ \ \ \ \ \ \ \ \ \ \ \ \ \ \ \ \ \ \ \ \ \ \ \ \ \ \ \ \ \ \ \ \ \ \ \times \left[ {\Gamma _\mu ^ +  \left( {j\left| {\mu ^ -  } \right.} \right)p_{st} \left( {j\left| {\mu ^ -  } \right.} \right) - \Gamma _\mu ^ -  \left( {j\left| {\mu ^ +  } \right.} \right)p_{st} \left( {j\left| {\mu ^ +  } \right.} \right)} \right]. \\
\end{array}
\end{eqnarray}
Here $S_{\nu \nu}$ and $S_{II}$ are the spectral densities of
current fluctuations (SDCF) for currents $I_{\nu}$ and $I$,
respectively; $A_{\nu} = e(I_{\nu}^+ + I_{\nu}^- )$ with
$I_{\nu}^{\pm} = e\sum_i p_{st}(i) \Gamma_{\nu}^{\pm}$; the
conditional probability $p(i \leftarrow j|\tau )$ for having state
$|i >$ at the time $t = \tau
> 0$ under the condition that the state was $|j >$ at an earlier
time $t = 0$ obeys the same master equation as for the probability
$p(i,t)$; the stationary probability $p_{st}(i)$ is
defined as $p(i \leftarrow j|\tau  \to \infty ) = p_{st}(i) \delta
_{ij}$; $\hat B = Re \{( i\omega \hat 1 - \hat M )^{-1} \}$;
$\left\langle {j|\nu ^{\pm }} \right\rangle$ is the state obtained
from the state $|j> = (n_j ,m_j )$ by transferring an electron
across the $\nu$-junction to the right(+)/left(-).

For the structure under study, using the well-known expression of
the tunneling rate across a junction (see, eqs.(4,5) in [9]), in
principle, one can calculate the SDCFs (1). In practice, however,
one can not solve the master equation exactly with all possible
states. Recently [9], we have shown that at zero temperature and
in the first Coulomb staircase region, $V_{s1} \leq  V  \leq
V_{s2}$, where $V_{s2} = e/2C_r $ and $V_{s1}$ is the maximum from
$e/2C_l$ and $e|C_l - C_m |/2C_r (C_l + C_m )$ (assuming $C_l \geq
C_r$), the master equation can be exactly solved and therefore the
I-V characteristics can be derived for the range of parameters as
\begin{equation}
C_r \leq C_l \leq 3C_r \ {\rm{and}} \ C_m \leq C_r (3C_r - C_l
)/(C_l - C_r ).
\end{equation}
Under this condition all probabilities $p_{st}(i)$ are equal to zero
except those for three states $|1> = (0,0), |2> = (1,0) \
{\rm{and}} \ |3> = (1,-1)$: $p_{st}(1) = bc/ \Sigma_{\Gamma} ; \
p_{st}(2) = ca/ \Sigma_{\Gamma}$ and $ \ p_{st}(3) = ab/
\Sigma_{\Gamma}$, where
\begin{eqnarray}
\begin{array}{l}
 a \equiv \Gamma \left( {2 \leftarrow 1} \right) = \left[ {C_l\left( {C_m  + C_r} \right)/e\Sigma_c R_l } \right]\left( {V - e/2C_l } \right), \\
 b \equiv \Gamma \left( {3 \leftarrow 2} \right) = \left[ {C_r\left( {C_m  + C_l} \right)/e\Sigma_c R_r } \right]
 \left( {V - e\left( {C_l  - C_m } \right)/2C_r \left( {C_l  + C_m } \right)} \right), \\
 c \equiv \Gamma \left( {1 \leftarrow 3} \right) = \left[ {C_lC_r /e\Sigma_c R_m } \right]\left( {e/2C_l  + e/2C_r  - V}
 \right),
\end{array}
\end{eqnarray}
$\Sigma_{\Gamma} = ab + bc + ca$. With (3) the SDCFs (1) become
\begin{eqnarray}
\begin{array}{l}
 S_{ll} = 2eI\left( {1 + 2aB_{12} } \right),\ S_{mm} = 2eI\left( {1 + 2aB_{31} } \right),
 \ S_{rr} = 2eI\left( {1 + 2aB_{23} } \right), \\
 S_{II} = 2eI\sum\limits_\nu  {g_\nu ^2 }  + 4eI[g_l a\left( {g_m B_{11}  + g_l B_{12}  + g_r B_{13} } \right)
 + g_r b\left( {g_m B_{21}  + g_l B_{22}  + g_r B_{23} } \right) \\
 \ \ \ \ \ \ \ \ \ \ \ \ \ \ \ \ \ \ \ \ \ \ \ \ \ + \ g_m c\left( {g_m B_{31}  + g_l B_{32}  + g_r B_{33} } \right).
\end{array}
\end{eqnarray}
The matrix $(i\omega \hat{I} - \hat{M})$ has then a simple form
giving the matrix $\hat{B}$ with elements
\begin{equation}
 B_{ij} = Re\{D_{ij}[i\omega (ab + bc + ca - \omega^2 ) - \omega^2 (a + b +
 c)]^{-1}\},
\end{equation}
where $D_{11} = bc - \omega^2 + i\omega (b+c); \ D_{12} = bc; \
D_{13} = bc + i\omega c ; \ D_{21} = ac + i\omega a; \ D_{22} = ac
- \omega^2 + i\omega (a+c); \ D_{23} = ac; \ D_{31} = ab; \ D_{32}
= ab + i\omega b; \ D_{33} = ab - \omega^2 + i\omega (a+b)$. The
expression (4) (with (3) and (5)) is our main analytical result obtained for
the first Coulomb staircase region in the condition (2).
Substituting (3) and (5) into (4) we see that for the model under
study all $S_{\nu \nu}(\omega)$ ($\nu = l, m, r$) are identical
\begin{eqnarray}
\frac{{S_{\nu \nu } \left( \omega  \right)}}{{2eI}} = 1 -
\frac{{2abc\left( {a + b + c} \right)}}{{\left( {ab + bc + ca -
\omega ^2 } \right)^2  + \omega ^2 \left( {a + b + c} \right)^2
}}.
\end{eqnarray}
In the limit of zero frequency the noises $S_{\nu \nu}(0)$ and
$S_{II}(0)$ are coincident with a single Fano factor
\begin{equation}
F = [1/a^2 + 1/b^2 + 1/c^2][1/a + 1/b + 1/c]^{-2}.
\end{equation}
In the opposite limit of large frequency $F_{\nu \nu} \rightarrow
1$ (the Poissonian value), whereas $F_{II} \rightarrow \sum_{\nu}
g_{\nu}^2 < 1 $.

For given values of device parameters (capacitances and tunneling
resistances) satisfying the condition (2) it is easy to calculate
the SDCFs (4) in a large range of frequency. To this end we use
the zero temperature tunneling rate across a junction $\Gamma =
\Theta (-\Delta F)|\Delta F|/e^2 R_t$, where $R_t$ is the junction
tunneling resistance, $\Delta F$ is the change in the free energy
$F$ of the system after the tunneling event has occurred. For the
model under study $F(i) = (en_i - C_l V/2)^2 /2C_l^* + (em_i + C_r
V/2)^2 /2C_r^* + (en_i - C_l V/2)(em_i + C_r V/2)/2C_m^* + eV(n_l
- n_r )/2 - (C_l + C_r )V^2 /8$, where $C_l^* = \Sigma_C /(C_l +
C_m ), C_r^* = \Sigma_C /(C_l + C_m ), C_m^* = \Sigma_C /C_m$ ;
$n_l (n_r)$ is the number of electrons that have entered the
structure from the left (right) [9]. In calculations as well as in
MC-simulations below the elementary charge $e$, the capacitance
$C_r$ and the tunneling resistance $R_r$ are chosen as the basic
units. The voltage, the current and the frequency are then
measured in the units of $e/C_r, \ e/C_r R_r$ and $(C_r R_r
)^{-1}$, respectively. Fig.1$b$ shows the normalized SDCFs $S_{\nu
\nu}(\omega )/2eI$ (two upper lines) and $S_{II}(\omega )/2eI$
(two lower lines) calculated at the same bias voltage $V = 0.5$
for two cases of I-V characteristics with an PDC (dashed line) and
an NDC (solid line) as shown correspondingly in Fig.1$a$. This
result should be in comparison with that for the SET shown in
Fig.3 of [7]. For the more interesting case of NDC the Fano factor
is plotted versus the bias voltage in the inset of this figure.
Clearly, the noise is considerably greater in the NDC regime, the
factor $F$ is however still limited by the Poissonian value. An
enhancement of the SN in an NDC regime is observed in various
structures due to a strong fluctuation of current. The present
result of $F \leq 1$ is different from that observed in resonant
tunneling diode devices where the electrostatic potential
fluctuation of the band bottom leads to a super-Poissonian noise
[1,4]. It should be emphasized that the origin of the
super-Poissonian noise is related to the nature of potential
fluctuations producing a positive feedback of charge which is
absent in the MDQDS under study. On the other hand, the
sub-Poissonian noise accompanied by an NDC has been observed in
$GaAs-AlAs-GaAs$ tunneling structure with embedded self-assembled
$InAs$ quantum dots in the single-electron tunneling regime [11]
and also suggested in strongly correlated double quantum dot
systems in the Kondo regime [12].

For a large range of bias voltage we simulate the noise using
basically the MC-program [9] which was shown to give I-V curves in
good agreement with analytical calculations (see Fig.2 in [9]).
However, it is impossible to simulate the spectral density in the
limit of zero frequency and we have to consider the low-frequency
limit $\omega_c = 10^{-3}$ (indicated by the arrow in Fig.1$b$). For
all frequencies $\omega \ge \omega_c$ the simulation noises are
practically coincident with corresponding analytical curves in
Fig.1$b$. As an additional test, our noise program has well
reproduced the qualitative behavior of experimental data in Fig.1 of
[2] for a single dot structure. Fig.2 presents the voltage
dependence of normalized simulation SDCFs for sample with parameters
given in the figure. The calculation has been performed at $\omega =
\omega_c$ in a range of $V$ where the I-V curve (dashed line)
exhibits several peaks and valleys. Clearly, two solid curves
describing $S_{\nu \nu}(V)$ and $S_{II}(V)$ have practically the
same form, though at the chosen finite frequency there is still a
considerable separation between them. At each bias voltage the Fano
factor $F$ is somewhere between these normalized noises and we can
guess that the $F(V)$-dependence should have the same saw-tooth
behavior like solid curves in Fig.2. The most interesting feature
observed in this figure is the modulation of the noise amplitude as
a function of $V$ with peaks at the points where the NDC reaches the
highest magnitude and moreover the noise seems to be always
sub-Poissonian even in NDC regimes. Such a suppression of noise is
due to the strong Coulomb interaction as generally suggested in
ref.[8]. We like to mention that a similar voltage-dependent
behavior of the Fano factor has been experimentally observed in [11]
for a single quantum dot structure. The gradual decrease of noise
peaks at NDC regimes as the voltage increases is closely related to
the corresponding decrease of the peak-to-valley ratio of current as
can be seen in Fig.2. Note that qualitatively all the simulation
results discussed are not particular for the chosen frequency
$\omega_c$. Thus, both the analytical results of eqs.(6) and (7) and
the MC-simulation data suggest that the SN in the MDQDS under study
is always sub-Poissonian though it is considerably enhanced in NDC
regimes. This is the main conclusion of the present work which, as
shown by additional simulation data (should be published elsewhere),
is well preserved even in the case gates are included.

One of authors (P.D.) acknowledges the CNRS for financial support under PICS No.404.
The work in Hanoi was supported in part by the Natural Science Council of Vietnam
and by the VAST research grant on simulation of nano-devices.

\newpage

\textbf{FIGURE CAPTIONS}

 \vspace*{1cm} FIG.1 The I-V characteristics (a) and corresponding normalized
 SDCFs calculated at V = 0.5 (b) for $R_l = 1.1$ (PDC - dashed lines) and
 $R_l = 0.2$ (NDC - solid lines), everywhere $C_l = 1.5, C_m = 2.0, R_m =
 2.0$. Insets: in (a): circuit diagram of the model, in (b): the Fano factor
 as a function of the bias for the NDC case.

 \vspace*{1cm} FIG.2 The voltage dependence of normalized SDCFs: solid curves (upper
 for $S_{\nu \nu }$ and lower for $S_{II}$) at the frequency $\omega_c = 10^{-3}$.
 The corresponding I-V characteristics is shown by the dashed curve.
 Simulation parameters: $C_l = 1.0, C_m = 1.0, R_l = 1.0, R_m = 10.0$,
 zero temperature.


\begin{thebibliography}{99}
\bibitem{10fnoise}  The expression used for $\langle I(t)\rangle$
                    is simply derived from the work done by the
                    voltage source considering the individual
                    currents $I_\nu (t)$
\end{thebibliography}
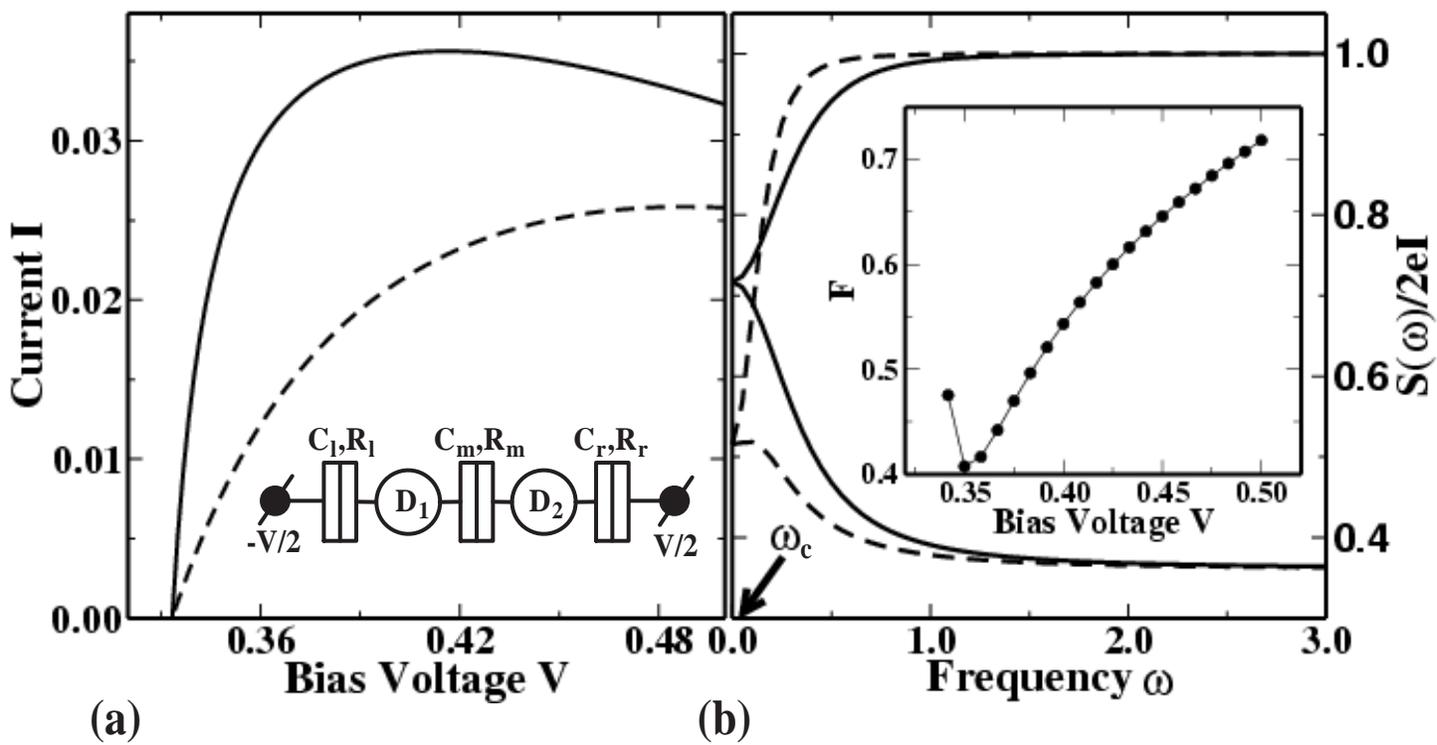
\end{document}